\documentclass[11 pt,oneside,onecolumn,a4paper]{article}
\usepackage{amsmath}
\usepackage{graphicx}
\DeclareGraphicsExtensions{.eps,.ps,.pdf}
\usepackage{bbm}
\usepackage{bm}% bold math
\usepackage{epsfig}
\usepackage[T1]{fontenc}
\usepackage{esint}
\usepackage{amssymb}

\usepackage{lipsum}
mm \leftmargin 5pt \rightmargin 5pt \hoffset= -15mm

\title{Time dilation in the oscillating decay laws of moving two-mass unstable quantum states
}
\author{Filippo Giraldi}

\date{\small{School of Chemistry and Physics, University of KwaZulu-Natal\\ 
and National Institute for Theoretical Physics (NITheP)\\
Westville Campus, Durban 4000, South Africa
\vspace{1em}\\Gruppo Nazionale per la
Fisica Matematica (GNFM-INdAM)\\
c/o Istituto Nazionale di Alta Matematica Francesco Severi\\
Citt\`{a} Universitaria, Piazza Aldo Moro 5, 00185 Roma, Italy}}

\begin{document}

\maketitle

\def\bbm[#1]{\mbox{\boldmath$#1$}}

\vspace{0em}

%%%%%%%%%%%%%%%%%%%%% Publisher's Area please ignore %%%%%%%%%%%%%%
%\catchline{}{}{}{}{}
%%%%%%%%%%%%%%%%%%%%%%%%%%%%%%%%%%%%%%%%%%%%%%%%%%%%%%%%%%%%%%%%%%%

%\maketitle

%\begin{history}
%%\received{Day Month Year}
%\revised{Day Month Year}
%\accepted{Day Month Year}
%\comby{(xxxxxxxxxx)}
%\end{history}

\vspace{0em}

\begin{abstract}
The decay of a moving system is studied in case the system is initially prepared in a two-mass unstable quantum state. The survival probability $\mathcal{P}_p(t)$ is evaluated over short and long times in the reference frame where the unstable system moves with constant linear momentum $p$. The mass distribution densities of the two mass states are tailored as power laws with powers $\alpha_1$ and $\alpha_2$ near the non-vanishing lower bounds $\mu_{0,1}$ and $\mu_{0,2}$ of the mass spectra, respectively. If the powers $\alpha_1$ and $\alpha_2$ differ, the long-time survival probability $\mathcal{P}_p(t)$ exhibits a dominant inverse-power-law decay and is approximately related to the survival probability at rest $\mathcal{P}_0(t)$ by a time dilation. The corresponding scaling factor $\chi_{p,k}$ reads $\sqrt{1+p^2/\mu_{0,k}^2}$, the power $\alpha_k$ being the lower of the powers $\alpha_1$ and $\alpha_2$. If the two powers coincide and the lower bounds $\mu_{0,1}$ and $\mu_{0,2}$ differ, the scaling relation is lost and damped oscillations of the survival probability $\mathcal{P}_p(t)$ appear over long times. 
By changing reference frame, the period $T_0$ of the oscillations at rest transforms in the longer period $T_p$ according to a factor which is the weighted mean of the scaling factors of each mass, with non-normalized weights $\mu_{0,1}$ and $\mu_{0,2}$.
\end{abstract}

\maketitle
%\keywords{information backflow, recoherence, structured reservoirs, spectral gaps}

%\tableofcontents  % optional

%\markboth{F. Giraldi}{Sequences of information backflow in local dephasing channels with spectral gaps}

\section{Introduction}\label{1}
\vspace{-0em}

In many experiments of high-energy physics or astrophysical phenomena, the involved unstable particles move in the laboratory frame of the observer at relativistic or ultrarelativistic velocities. For this reason, a theoretical description of the relativistic transformations of the decay laws of moving unstable quantum systems is fundamental. The literature which concerns this subject is vast. See Refs. \cite{Khalfin,BakamjianPR1961RQT,ExnerPRD1983,HEP_Stef1996,TD_StefanovichXivHep2006,GiacosaFP2012,UrbPLB2014,UrbAPB2017,GiacosaRelDecayXiv2018}, to name but a few. In the rest reference frame of the moving unstable system the 
nondecay or survival probability is expressed in term of the mass distribution density (MDD). The low-mass, integral and analytic properties of the MDD determine the long-time decay of the survival probability \cite{FondaGirardiRiminiRPP1978,UrbanowskiEPJD2009,UrbanowskiCEJP2009,GEPJD2015}. In the reference frame where the unstable system moves with constant linear momentum, the survival probability has been evaluated in various ways by adopting quantum theory and special relativity \cite{HEP_Stef1996,HEP_Shir2004,HEP_Shir2006,TD_StefanovichXivHep2006,GiacosaAPPB2016,GiacosaAPPB2017,UrbAPB2017,GiacosaRelDecayXiv2018}. The transformations of the decay laws of a moving unstable quantum system, which are induced by the change of reference frame, raise questions on the appearance of the relativistic dilation of times in quantum decays. As matter of fact, relativistic time dilation is considered to appear over the short-time exponential decay of the survival probability. The literature which concerns this subject is vast. See Refs. \cite{HEP_Stef1996,HEP_Shir2004,TD_StefanovichXivHep2006,HEP_Shir2006,GiacosaAPPB2016,GiacosaAPPB2017,UrbAPB2017}, to name but a few.

Recently, the long-time behavior of the survival probability has been analyzed for arbitrary values of the constant linear momentum of a moving unstable quantum system and for MDDs which 
are tailored as power laws near the (non-vanishing) lower bound of the mass spectrum \cite{Gxiv2018}. The decay laws which are detected in the rest reference frame of the unstable system are reproduced by the condition of vanishing linear momentum, while non-vanishing values of the linear momentum refer to the laboratory frame of an observer where the unstable particle is moving. Over long times, the survival probability at rest transforms in the reference frame where the unstable system moves with constant linear momentum, according to a time dilation. The scaling factor of the time dilation is determined by the lower bound of the mass spectrum and by the linear momentum. Also, the scaling factor results to be the ratio of the asymptotic value of the instantaneous mass \cite{UrbPLB2014,UrbAPB2017,UrbanowskiEPJD2009,UrbanowskiCEJP2009,UrbanowskiPRA1994} and of the instantaneous mass at rest of the unstable system. By considering the asymptotic value of the instantaneous mass as the effective mass of the unstable system over long times, the scaling factor coincides with the relativistic Lorentz factor of the moving system. This property allows to interpret the time dilation which appears in the long-time decay laws, in terms of the relativistic time dilation. See Ref. \cite{Gxiv2018} for details.

Oscillating decay laws have always attracted a great deal of attention in quantum theory. Oscillating behaviors are obtained in the exponential regime of the decay of an unstable state in the referential $N$-level Friedrichs model \cite{OscFM}. Oscillatory behaviors of the decay laws are obtained if the energy distribution density of the quantum system deviates from the Breit-Wigner form \cite{GPoscillatingDecaysQM2012}. Referential examples of oscillating decays are found in neutrino and unstable meson systems. The literature which concerns this subject is vast. See Refs. \cite{Perkins,HEP_Shir2004,HEP_Shir2006,ShirokovNaumov2006} to name but a few. Oscillating decay laws are obtained in theoretical models which involve unstable two-mass states \cite{HEP_Shir2004,HEP_Shir2006,ShirokovNaumov2006}. 
%The terms which describe the survival amplitude interfere and oscillations appear in the survival probability. 
In Ref. \cite{HEP_Shir2006} the initial two-mass state is considered to be a superposition of two eigenstates of the Hamiltonian with different eigenvalues. In this condition the survival probability exhibits regular undamped oscillations. In the reference frame where the unstable system moves with constant linear momentum, the survival probability is obtained from the survival probability at rest via a time dilation which is not Einsteinian if the two mass eigenstates differ. See Ref. \cite{HEP_Shir2006} for details. In Ref. \cite{HEP_Shir2004}, the initial two-mass state is the superposition of two ustable states which are described by MDDs of Breit-Wigner form. The survival probability exhibits oscillations which are enveloped in an exponential decay. The relativistic dilatation of times holds if the two rest masses are approximately equal. See Ref. \cite{HEP_Shir2004} for details.

As a continuation of the scenario described above, here, we consider a moving system which is initially prepared in a two-mass unstable quantum state. The MDDs which describe the two mass states are tailored as power laws near the lower bound of the corresponding mass spectrum. In light of the study which is performed in Ref. \cite{HEP_Shir2004}, we expect damped oscillations of the survival probability in case the two MDDs are somehow different from each other. 
We intend to study the decay laws in a reference frame where the two-mass unstable quantum state is an eigenstate of the linear momentum with a nonvanishing eigenvalue. We aim to compare the corresponding decay laws with the decay laws which are detected in the rest reference frame of the moving unstable system. 
We intend to study the transformations of the survival probability and search for possible time dilation. We also aim to describe how the eventual oscillations of the survival probability transform due to the change of reference frame.

The paper is organized as follows. Section \ref{2} is devoted to the description of the decay of general moving single-mass unstable quantum systems in term of the MDD. In Section \ref{3}, the decay laws of moving two-mass unstable quantum states are evaluated over short and long times. Section \ref{4} is devoted to the transformations of the decay laws which are due to the change of reference frame, and to the appearance of time dilation. Summary and conclusion are reported in Section \ref{5}. Demonstrations of the results are provided in Appendix.

\section{Moving single-mass unstable quantum systems}\label{2}

For the sake of clarity and convenience, we report below some details on the transformations of the survival probability, which are due to the change of reference frame, by following Ref. \cite{UrbAPB2017}. Let the quantum state of the unstable particle belong to the Hilbert space $\mathcal{H}$, and let the state kets $|m,p\rangle$ represent the common eigenstates of the linear momentum $P$ operator with eigenvalue $p$, and of the Hamiltonian $H$ self-adjoint operator, with eigenvalue $m$. These assumptions mean that the following equalities, $P|m,p\rangle =p |m,p\rangle$ and $H|m,p\rangle =E(m,p) |m,p\rangle$, hold over the mass spectrum, and for every value of the linear momentum $p$. The spectrum of the Hamiltonian is assumed to be continuous with lower bound $\mu_0$.

Let the state ket $|\phi\rangle$ belong to the Hilbert space $\mathcal{H}$ and represent the initial unstable  state of the quantum system. This state can be expressed via the eigenstates $|m,0\rangle$ of the Hamiltonian as $|\phi\rangle=\int_{\mu_0}^{\infty} \langle 0,m||\phi\rangle |m,0\rangle dm$. In the present notation, $\langle 0,m|$ and $\langle p,m|$ are the bras of the state kets $|m,0\rangle$ and $|m,p\rangle$, respectively. In the rest reference frame of the unstable system the survival amplitude reads $A_0(t)=\langle \phi| e^{-\imath H t} |\phi\rangle$, where $\imath$ 
is the imaginary unit, and is given by the following integral expression \cite{UrbAPB2017,UrbPLB2014,UrbanowskiEPJD2009,UrbanowskiCEJP2009,FondaGirardiRiminiRPP1978},
\begin{eqnarray}
A_0(t)=\int_{\mu_0}^{\infty} \omega\left(m\right) e^{-\imath m t} dm. \label{A0Int}
\end{eqnarray}
 The function $\omega\left(m\right)$ is the MDD of the unstable system and reads $\omega\left(m\right)=\left|\langle 0,m||\phi\rangle\right|^2$. The MDD is determined by the initial state and by the Hamiltonian of the system via the eigenstates $|m,0\rangle$. In the rest reference frame of the moving unstable system, the probability $\mathcal{P}_0(t)$ that the decaying system remains in the initial state at the time $t$, is referred as the survival probability at rest and reads $\mathcal{P}_0(t)=\left|A_0(t)\right|^2$.

In the reference frame where the system moves with constant linear momentum $p$, the transformed survival amplitude $A_p(t)$ is given by the expression $\langle \phi_p|e^{-\imath H t}|\phi_p\rangle$. The state $|\phi_p\rangle$ is an eigenstate of the linear momentum $P$, with eigenvalue $p$, and
represents the state $|\phi \rangle$ in the reference frame where the system moves with linear momentum $p$. By adopting this notation, the state $|\phi_{0} \rangle$ represents the initial unstable state $|\phi\rangle$ of the system in the rest reference frame. The transformed survival amplitude $A_p(t)$ is obtained in various ways \cite{HEP_Stef1996,HEP_Shir2004,HEP_Shir2006,GiacosaAPPB2016,GiacosaAPPB2017,UrbAPB2017,GiacosaRelDecayXiv2018}, and results in the following fundamental integral form,
\begin{eqnarray}
A_p(t)=\int_{\mu_0}^{\infty} \omega\left(m \right)
e^{-\imath \sqrt{ p^2+m^2}t} d m.
\label{Aptdef}
\end{eqnarray}
The survival probability $\mathcal{P}_p(t)$ is given by the square modulus of the above expression, $\mathcal{P}_p(t)=\left|A_p(t)\right|^2$. See Refs. \cite{HEP_Stef1996,HEP_Shir2004,HEP_Shir2006,GiacosaAPPB2016,GiacosaAPPB2017,UrbAPB2017,GiacosaRelDecayXiv2018} for details.

\subsection{Mass distribution densities}\label{21}

Unstable quantum states are usually described by MDDs which consist in a Breit-Wigner form plus, eventually, a form factor and an additional term which provides a low-mass power-law profile \cite{FondaGirardiRiminiRPP1978,UrbPLB2014,UrbAPB2017,UrbanowskiEPJD2009,UrbanowskiCEJP2009,HEP_Shir2004,GiacosaFP2012}. 
In Ref. \cite{Gxiv2018}, the survival amplitude $A_p(t)$, the survival probability $\mathcal{P}_p(t)$, the instantaneous mass $M_p(t)$ and the instantaneous decay rate $\Gamma_p(t)$ are evaluated, over short and long times, for an arbitrary value $p$ of the constant linear momentum of the moving unstable quantum system, in case 
the MDD is tailored as a nonegative power law near the lower bound $\mu_0$ of the mass spectrum. The resulting survival probability $\mathcal{P}_p(t)$ decays as an inverse power law over long times and is related to the survival probability at rest $\mathcal{P}_0(t)$ by a scaling law. For the sake of clarity and convenience, the MDDs under study and the main results of Ref. \cite{Gxiv2018} are reported below.

The MDD of the system is described by the auxiliary function $\Omega\left(\xi\right)$, over the infinite support $\left[\right.\mu_0,\infty\left.\right)$, via the following scaling law, $\omega\left(m_s \xi \right)= \Omega\left(\xi\right)/m_s $. The scaling relation holds for every $\xi\geq \xi_0$. The parameter $m_s$ represents a general scale mass. The dimensionless variable $\xi$ is defined as $\xi=m/m_s$, and the parameter $\xi_0$ reads $\xi_0=\mu_0/m_s$. The condition of non-vanishing lower bound of the mass spectrum is equivalent to the constraint $\xi_0>0$. The power-law behavior of the MDD is given by the relation
\begin{eqnarray}
\Omega\left(\xi \right)= \left(\xi-\xi_0\right)^{\alpha}
\Omega_0\left(\xi \right),
\label{Omegaalpha}
\end{eqnarray}
with $\alpha\geq 0$ and $\Omega_0  \left(\xi_0 \right)>0$.
 The function $\Omega_0\left(\xi \right)$ and the derivatives $\Omega^{(j)}_0\left(\xi \right)$ must be summable, for every $j=1, \ldots, \lfloor \alpha \rfloor +4$, and continuously differentiable in $\left[\mu_0,+\infty\right.\left.\right)$, for every $j=1, \ldots, \lfloor \alpha \rfloor +3$. Consequently, the limits $\lim_{\xi\to\xi_0^+}\Omega^{(j)}_0\left(\xi \right)$ exist finite and coincide with $\Omega^{(j)}_0\left(\xi_0 \right)$ for every $j=0, \ldots, \lfloor \alpha \rfloor+4$. The functions $\Omega_0^{(j)}\left(\xi \right)$ are required to decay sufficiently fast as $\xi\to+\infty$, so that the auxiliary function $\Omega\left(\xi\right)$ and the derivatives $\Omega^{(j)}\left(\xi\right)$ vanish as $\xi\to+\infty$, for every $j=0, \ldots, \lfloor \alpha \rfloor$.

\subsection{Time dilation in the decay laws of moving single-mass unstable quantum states
}\label{22}
The survival amplitude $A_p(t)$, given by the integral expression (\ref{Aptdef}), is determined by the auxiliary function $\Omega\left(\xi \right)$ via the following form \cite{Gxiv2018},
\begin{eqnarray}
A_p(t)=\int_{\xi_0}^{\infty} \Omega\left(\xi \right)
e^{-\imath \eta \tau} d \xi,
\label{Aptxi}
\end{eqnarray}
where $\tau=m_s t$, $\eta=\sqrt{\rho^2+\xi^2}$ and $\rho=p/m_s$. The short-time behavior of the survival amplitude is obtained from the asymptotic analysis of the above integral form and leads to an approximate algebraic decay of the survival probability $\mathcal{P}_p(t)$ over short times,
\begin{eqnarray}
\mathcal{P}_p(t)\sim 1 - \pi_0 t^2, \label{Pptshort}
\end{eqnarray}
for $t \ll 1/m_s$. The coefficient $\pi_0$ is given in the Appendix. The long-time behavior of the survival probability is approximated by the form below,
\begin{eqnarray}
\mathcal{P}_p(t)\sim c^2_0 \left(\frac{\chi_p}{m_s t}\right)^{2\left( 1+\alpha\right)},
\label{Pplongt}
\end{eqnarray}
for $t \gg 1/m_s$. The constant $c_0$ is defined in the Appendix. The factor $\chi_p$ is determined by the linear momentum $p$ and by the lower bound $\mu_0$ of the mass spectrum,
\begin{eqnarray}
\chi_p = \sqrt{1+\frac{p^2}{\mu_0^2}}. \label{Chip}
\end{eqnarray}
The long-time survival probability at rest $\mathcal{P}_0(t)$ is obtained from Eq. (\ref{Pplongt}) for vanishing value of the linear momentum $p$, and reads
\begin{eqnarray}
\mathcal{P}_0(t)\sim c^2_0 \left(m_s t\right)^{-2\left( 1+\alpha\right)},
\label{P0longt}
\end{eqnarray}
for $t \gg 1/ m_s$.

For the MDDs under study, the survival probability $\mathcal{P}_p\left(t\right)$ is related to the survival probability at rest $\mathcal{P}_0\left(t\right)$, approximately over long times, by the following scaling law,
\begin{eqnarray}
\mathcal{P}_p\left(t\right) \sim \mathcal{P}_0\left(\frac{t}{ \chi_p}\right), \label{PpP0L}
\end{eqnarray}
for $t \gg 1/m_s$. The above approximate long-time scaling relation is obtained by comparing Eqs. (\ref{Pplongt}) and (\ref{P0longt}). This scaling relation describes the transformation of the survival probability at rest, over long times, in the reference frame where the system moves with constant linear momentum $p$. The transformation consists in an approximate time dilation which is determined by the scaling factor 
$\chi_p$. This factor coincides with the ratio of the asymptotic value of the instantaneous mass $M_p\left(\infty\right)$ and of the instantaneous mass at rest $M_0\left(\infty\right)$ of the moving unstable system, $\chi_p=M_p\left(\infty\right)/M_0\left(\infty\right)$, where $M_p\left(\infty\right)=\sqrt{\mu_0^2+p^2}$, and 
$M_0\left(\infty\right)=\mu_0$. The equality $\chi_p=\sqrt{1+p^2/\mu_0^2}$, suggests that the scaling factor coincides with the relativistic Lorentz factor of a mass at rest $\mu_0$ which moves with linear momentum $p$ or, equivalently, with constant velocity 
$1\Big/\sqrt{1+\mu_0^2\big/p^2}$. If the asymptotic value of the instantaneous mass is considered as the effective long-time mass of the unstable system, the time dilation of the survival probability coincides with the relativistic time dilation of the instantaneous mass at rest $M_0\left(\infty\right)$ which moves with constant linear momentum $p$. In this way, the time dilation which is found in the transformation of the decay laws, can be interpreted via the theory of special relativity. See Ref. \cite{Gxiv2018} for details.

\section{Moving two-mass unstable quantum states}\label{3}

At this stage, we study the decay laws of moving two-mass unstable quantum states
 \cite{HEP_Shir2004}. 
Let the unstable system be initially prepared in a pure state $|\phi\rangle$ which is the superposition of two unstable mass states $|\phi_1\rangle$ and $|\phi_2\rangle$, 
\begin{eqnarray}
|\phi\rangle= l_1|\phi_1\rangle+l_2|\phi_2\rangle.
\label{2Minitial}
\end{eqnarray}
 The two states $|\phi_1\rangle$ and $|\phi_2\rangle$ are consider to be approximately orthogonal, $\langle \phi_1\left|\right| \phi_2\rangle\simeq 0$. Consequently, the normalization condition, $\langle \phi||\phi\rangle=1$, requires that the coefficients $l_1$ and $l_2$ fulfill the following constraint, $\left|l_1\right|^{2}+\left|l_2\right|^{2}=1$.

 Let the function $\omega_j\left(m\right)$ be the MDD which describes the $j$th unstable mass state $|\phi_j\rangle$, for every $j=1,2$. This means that the following equality, $\omega_j\left(m\right)=\left|\langle 0,m| | \phi_j\rangle\right|^2$, holds over the support $\big[\mu_{0,j}, \infty\big)$. 
Let $\Omega_j\left(\xi\right)$ be the auxiliary function of the MDD $\omega_j\left(m\right)$ for every $j=1,2$. This means that the following relation, $\omega_j\left(m_s \xi  \right)=\Omega_j\left(\xi\right)/m_s$, holds for every $\xi\geq \xi_{0,j}$, and for every $j=1,2$, where $\xi_{0,j}=\mu_{0,j}/m_s>0$. The MDDs under study belong to the class which is defined in Section \ref{21}. Consequently, the low-mass power-law behavior is given by the relation
\begin{eqnarray}
\Omega_j\left(\xi \right)= \left(\xi-\xi_{0,j}\right)^{\alpha_j}
\Omega_{0,j}\left(\xi \right),
\label{Omegaalphaj}
\end{eqnarray}
where $\alpha_j\geq 0$, and $\Omega_{0,j}  \left(\xi_{0,j} \right)>0$, for every $j=1,2$.

 In addition to the approximate orthogonality, it is also required that the time evolution of the mass states $|\phi_1\rangle$ or $|\phi_2\rangle$ is approximately orthogonal to the states $|\phi_2\rangle$ or $|\phi_1\rangle$, respectively. This means that the relation
\begin{eqnarray}
\langle\phi_{p,j}\left|e^{-\imath H t}\right|\phi_{p,k} \rangle\simeq 0,
\label{ort1}
\end{eqnarray}
holds for every $t\geq 0$, for every $j,k=1,2$, such that $j \neq k$, and for every nonnegative real value of the linear momentum $p$. The state $|\phi_{p,l} \rangle$ is an eigenstate of the linear momentum $P$, with eigenvalue $p$, and represents the state $|\phi_{l} \rangle$ in the reference frame where the system moves with linear momentum $p$. In this notation, the state $|\phi_{0,l} \rangle$ represents the mass state $|\phi_l\rangle$, for every $l=1,2$. The assumption (\ref{ort1}) is motivated by the almost exact conservation of the CP-parity in the decay of unstable meson systems \cite{HEP_Shir2004}. In fact, the condition (\ref{ort1}) holds if the unstable states $|\phi_{1} \rangle$ and $|\phi_{2} \rangle$ have different CP-parities. The condition (\ref{ort1}) can be expresses in terms of the MDDs $\omega_{1}\left(m\right)$ and $\omega_{2}\left(m\right)$ and results in the relation below,
\begin{eqnarray}
\int_{\mu_{\rm M}}^{\infty}
\sqrt{\omega_1\left(m\right)\omega_2\left(m\right)}e^{-\imath\left( \varphi_k\left(m\right)-\varphi_j\left(m\right)+\sqrt{m^2+p^2}t\right)}dm
\simeq 0,
\label{ort2}
\end{eqnarray}
for every $t\geq 0$, for every $j,k=1,2$, such that $j \neq k$, and for every nonnegative real value of the linear momentum $p$. The lower extremum $\mu_{\rm M}$ of the integration is defined as the maximum between the lower bounds of the two mass spectra, $\mu_{\rm M}=\max \left\{\mu_{0,1},\mu_{0,2}\right\}$. The function $\varphi_l\left(m\right)$ is defined over the support $\big[\mu_{0,l}, \infty\big)$ as the phase of the amplitude $\langle 0,m||\phi_l\rangle$, for every $l=1,2$. This means that the following relation, $\langle 0,m||\phi_l\rangle= \sqrt{\omega_l\left(m\right)}e^{- \imath \varphi_l\left(m\right)}$, holds over the support $\big[\mu_{0,l}, \infty\big)$, for every $l=1,2$. Qualitatively, the MDDs $\omega_1\left(m\right)$ and $\omega_2\left(m\right)$ are appreciably nonvanishing over different intervals, or the function $e^{-\imath\left( \varphi_1\left(m\right)-\varphi_2\left(m\right)\right)}$ oscillates fast enough over the open interval $\big[\mu_{\rm M}, \infty\big)$, that the integral (\ref{ort2}) is approximately vanishing for every $t \geq 0$, and for every nonvanishing value of the linear momentum $p$.

%At this stage, 
We are now equipped to evaluate the survival amplitude $A_p(t)$ in case the moving unstable system is initially prepared in the two-mass state (\ref{2Minitial}). Let the MDDs $\omega_1\left(m\right)$ and $\omega_2\left(m\right)$ belong to the class which is defined in Section \ref{21}. Let the constraint (\ref{ort2}) or, equivalently, (\ref{ort1}), be fulfilled for every $j,k=1,2$, such that $j\neq k$, for every $t \geq 0$ and for every nonvanishing value of the linear momentum $p$. Under these conditions, the terms $\langle\phi_{p,1}\left|e^{-\imath H t}\right|\phi_{p,2} \rangle$ and $\langle\phi_{p,2}\left|e^{-\imath H t}\right|\phi_{p,1} \rangle$ are negligible with respect to the terms $\langle\phi_{p,1}\left|e^{-\imath H t}\right|\phi_{p,1} \rangle$ and \\$\langle\phi_{p,2}\left|e^{-\imath H t}\right|\phi_{p,2} \rangle$. Consequently, the survival amplitude $A_p(t)$, which is given by the expression $\langle\phi_{p}\left|e^{-\imath H t}\right|\phi_{p} \rangle$, results in the following form,
\begin{eqnarray}
A_p(t)=\sum_{j=1}^2 \left|l_j\right|^2 A_{p,j}(t), \label{Ap12}
\end{eqnarray}
where $A_{p,j}(t)=\langle\phi_{p,j}|e^{-\imath H t} 
|\phi_{p,j}\rangle$, for every $j=1,2$.

\begin{figure*}
 \includegraphics[width=0.6\textwidth]{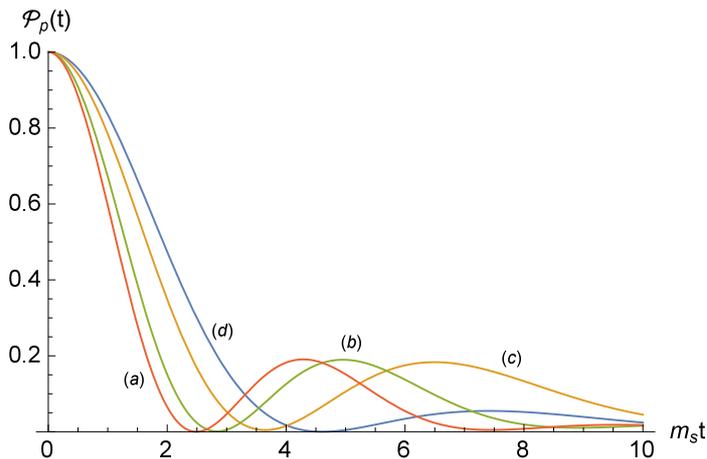}
\caption{(Color online) Survival probability $\mathcal{P}_p(t)$ versus $m_s t$, for 
$ 0 \leq m_s t \leq 10$, and different values of the linear momentum $p$. The computed initial two-mass states are described by Eq. (\ref{2Minitial}) with $l_1=l_2=1/\sqrt{2}$. The corresponding auxiliary functions are given by Eq. 
 (\ref{Ofigs}) with $\xi_{0,1}=1$ and $\xi_{0,2}=2$, and different values of the powers $\alpha_1$ and $\alpha_2$. Curve $(a)$ corresponds to $\alpha_1=0$, $\alpha_2=2$ and $p= 0 m_s$. Curve $(b)$ corresponds to $\alpha_1=0$, $\alpha_2=2$ and $p=  m_s$. Curve $(c)$ corresponds to $\alpha_1=0$, $\alpha_2=2$ and $p= 2 m_s$. Curve $(d)$ corresponds to $\alpha_1=1$, $\alpha_2=2$ and $p= 2 m_s$. 
}
\label{fig1}
\end{figure*}

\subsection{Survival probability}\label{31}

At this stage, we evaluate the survival probability of the moving two-mass unstable quantum system for MDDs which are defined in Section \ref{21} and fulfill the constraint (\ref{ort2}), or, equivalently, (\ref{ort1}). We expect long-time decay laws which depart from the single-mass decays \cite{Gxiv2018} if the MDDs are "`properly"' different from each other. The "`proper"' difference means that, at least, either the power-law profiles are different, $\alpha_1 \neq\alpha_2$, or the lower bounds of the mass spectra differ, $\mu_{0,1} \neq \mu_{0,2}$. We study the case $\alpha_1 \leq\alpha_2$, below. The decay laws which correspond to the condition $\alpha_1>\alpha_2$, are obtained from the case $\alpha_1<\alpha_2$, by exchanging the indexes $1$ and $2$. 

\begin{figure*}
\includegraphics[width=0.6\textwidth]{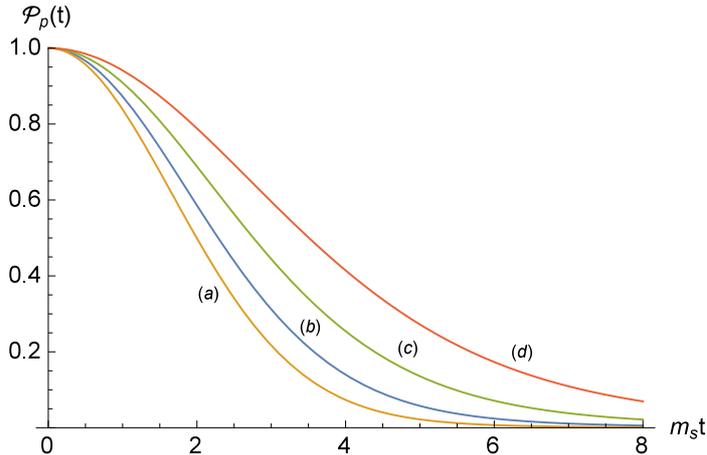}
\caption{(Color online) The survival probability $\mathcal{P}_p(t)$ versus $m_s t$, for 
$ 0 \leq m_s t \leq 8$, and different values of the linear momentum $p$. The computed initial two-mass states are described by Eq. (\ref{2Minitial}) with $l_1=l_2=1/\sqrt{2}$. The corresponding auxiliary functions are given by Eq. 
 (\ref{Ofigs}) for common values of the parameters $\xi_{0,1}$ and $\xi_{0,2}$ and different values of the powers, $\alpha_1=1$ and $\alpha_2=2$. Curve $(a)$ corresponds to $\xi_{0,1}=\xi_{0,2}=1$ and $p= 0 m_s$. Curve $(b)$ corresponds to $\xi_{0,1}=\xi_{0,2}=1$ and $p=  m_s$. Curve $(c)$ corresponds to $\xi_{0,1}=\xi_{0,2}=2$ and $p= 0 m_s$. Curve $(d)$ corresponds to $\xi_{0,1}=\xi_{0,2}=2$ and $p= 2 m_s$. 
}
\label{fig2}
\end{figure*}

 The short-time behavior of the survival probability is described by the following form,
\begin{eqnarray}
\mathcal{P}_p(t)\sim 1 - \bar{\pi}_0 t^2, \label{Pptshort2M}
\end{eqnarray}
for $t \ll 1/m_s$. The constant $\bar{\pi}_0$ is defined as $\bar{\pi}_0=2 \bar{a}_1-\bar{a}_0^2$. The constants $\bar{a}_0$ and $\bar{a}_1$ are given by the expressions below,
\begin{eqnarray}
&&\hspace{-3em}\bar{a}_0=\int_{ \mu_{\rm m}}^{\infty}\omega_{1,2}\left(m\right)\sqrt{m^2+p^2}  dm, \hspace{1em}
%nonumber \\&&\hspace{-3em}
\bar{a}_1=\frac{1}{2}\int_{ \mu_{\rm m}}^{\infty}\omega_{1,2}\left(m\right) \left(m^2+p^2\right)  dm.
\nonumber 
\end{eqnarray}
The lower extremum of integration $\mu_{\rm m}$ is defined as the minimum between the lower bounds of the two mass spectra, $\mu_{\rm m}=\min \left\{\mu_{0,1},\mu_{0,2}\right\}$.
 The function $\omega_{1,2}\left(m\right)$ is defined over the interval $\big[\mu_{\rm m}, \infty\big)$ as follows, $\omega_{1,2}\left(m\right)=\sum_{j=1}^2 \left|l_j\right|^2\omega_j\left(m\right)$. Each MDD vanishes outside the corresponding support.

\begin{figure*}
\includegraphics[width=0.6\textwidth]{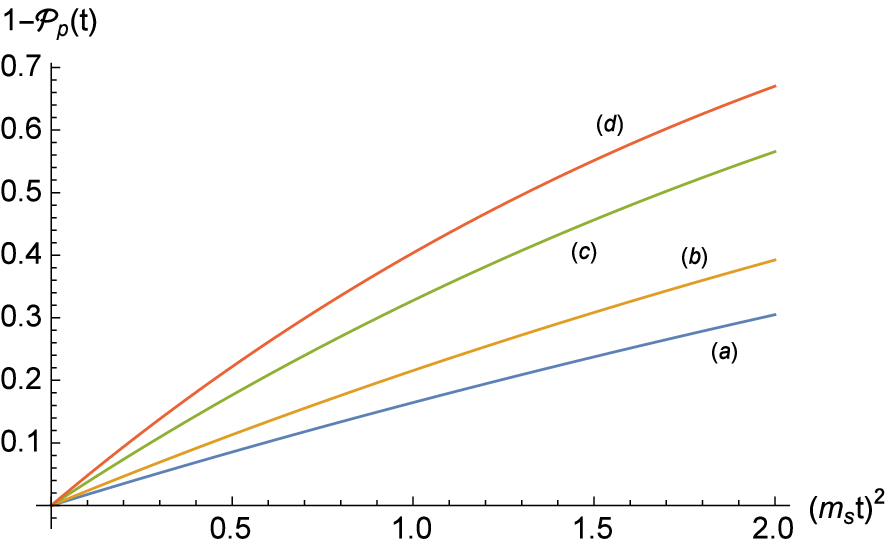}
\caption{(Color online) Quantity $\left(1-\mathcal{P}_p(t)\right)$ versus $\left(m_s t\right)^2$, for 
$ 0 \leq m_s t \leq 2$, and different values of the linear momentum $p$. The computed initial two-mass states are described by Eq. (\ref{2Minitial}) with $l_1=l_2=1/\sqrt{2}$. The corresponding auxiliary functions are given by Eq. 
 (\ref{Ofigs}) with $\xi_{0,1}=1$ and $\xi_{0,2}=2$, and different values of the powers $\alpha_1$ and $\alpha_2$. Curve $(a)$ corresponds to $\alpha_1=1$, $\alpha_2=2$ and $p= 2 m_s$. Curve $(b)$ corresponds to $\alpha_1=0$, $\alpha_2=2$ and $p= 2 m_s$. Curve $(c)$ corresponds to $\alpha_1=0$, $\alpha_2=2$ and $p= m_s$. Curve $(d)$ corresponds to $\alpha_1=0$, $\alpha_2=2$ and $p= 0 m_s$. The lines of different slope in agreement with the short-time behavior of Eq. (\ref{Pptshort2M}). 
}
\label{fig3}
\end{figure*}

As far as the long-time behavior is concerned, the survival probability results to be the sum of a dominant inverse-power-law and a dominant oscillating decay,
\begin{eqnarray}
\mathcal{P}_p(t)\sim \mathcal{P}^{({\rm p.l})}_p(t)+
\mathcal{P}^{({\rm osc})}_p(t), \label{P2MLt}
\end{eqnarray}
for $t \gg 1/\omega_s$. The scripts $({\rm p.l})$ stands for "`power law"', while 
the script $({\rm osc})$ means "`oscillating"'. If $\alpha_1<\alpha_2$, the dominant part of the long-time inverse-power-law decay $\mathcal{P}^{({\rm p.l})}_p(t)$ is 
\begin{eqnarray}
\mathcal{P}^{({\rm p.l.})}_p(t)\sim 
\mathfrak{P}^{({\rm p.l.})}_{p,1}\left(m_s t\right)^{-2\left(1+\alpha_1\right)},
\label{P2MLt1}
\end{eqnarray}
for $t \gg 1/m_s$. Instead, if $\alpha_1=\alpha_2$, the dominant part of the long-time inverse-power-law decay reads 
\begin{eqnarray}
\mathcal{P}^{({\rm p.l.})}_p(t)\sim \mathfrak{P}^{({\rm p.l.})}_{p,1,2}
\left(m_s t\right)^{-2\left(1+\alpha_1\right)}, 
\label{P2MLt12}
\end{eqnarray}
for $t \gg 1/m_s$. The coefficients $\mathfrak{P}^{({\rm p.l.})}_{p,1}$ and $\mathfrak{P}^{({\rm p.l.})}_{p,1,2}$ are given by the expressions below,
\begin{eqnarray}
&&\hspace{-3em}\mathfrak{P}^{({\rm p.l.})}_{p,1}=
\left|l_1 \right|^4 c_{0,1}^2 \chi_{p,1}^{2\left(1+\alpha_1\right)}, %\nonumber\\&&\hspace{-1em}
\hspace{1em}\mathfrak{P}^{({\rm p.l.})}_{p,1,2}=\sum_{j=1}^2
\left|l_j \right|^4 c_{0,j}^2 \chi_{p,j}^{2 \left(1+\alpha_j\right)}, \nonumber
\end{eqnarray}
where $c_{0,j}=\Gamma\left(1+\alpha_j\right)\Omega_{0,j}\left(\xi_{0,j}\right)$, for every $j=1,2$. 
The coefficients $\chi_{p,1}$ and $\chi_{p,2}$ are defined as follows,
\begin{eqnarray}
\chi_{p,j} = \sqrt{1+\frac{p^2}{\mu_{0,j}^2}}, \label{Chipj}
\end{eqnarray}
for every $j=1,2$. This notation refers to the scaling factor $\chi_p$ which is given by Eq. (\ref{Chip}). In Section \ref{22}, we have described how the scaling factor $\chi_p$ determines the time dilatation in the transformation of the survival probability for a moving single-mass unstable quantum system \cite{Gxiv2018}. Similarly, we show below how the parameters $\chi_{p,1}$ and $\chi_{p,2}$ influence the transformations of the terms $\mathcal{P}^{({\rm p.l.})}_p(t)$ and $\mathcal{P}^{({\rm osc})}_p(t)$ which are induced by the change of reference frame. 

\begin{figure*}
\includegraphics[width=0.6\textwidth]{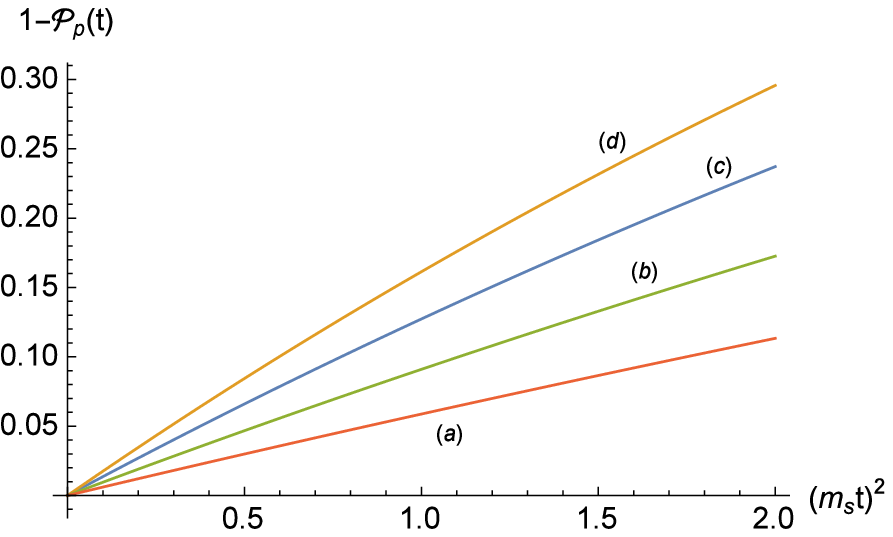}
\caption{(Color online) Quantity $\left(1-\mathcal{P}_p(t)\right)$ versus $\left(m_s t\right)^2$, for 
$ 0 \leq m_s t \leq 2$, and different values of the linear momentum $p$. The computed initial two-mass states are described by Eq. (\ref{2Minitial}) with $l_1=l_2=1/\sqrt{2}$. The corresponding auxiliary functions are given by Eq. 
 (\ref{Ofigs}) for common values of the parameters $\xi_{0,1}$ and $\xi_{0,2}$ and different of the powers, $\alpha_1=1$ and $\alpha_2=2$. Curve $(a)$ corresponds to $\xi_{0,1}=\xi_{0,2}=2$ and $p= 2 m_s$. Curve $(b)$ corresponds to $\xi_{0,1}=\xi_{0,2}=2$ and $p= 0 m_s$. Curve $(c)$ corresponds to $\xi_{0,1}=\xi_{0,2}=1$ and $p=  m_s$. Curve $(d)$ corresponds to $\xi_{0,1}=\xi_{0,2}=1$ and $p= 0 m_s$. 
}
\label{fig4}
\end{figure*}

Damped oscillations of the survival probability appear over long times if the lower bounds of the two MDDs differ, $\mu_{0,1}\neq\mu_{0,2}$. Under this condition, the dominant term of the oscillatory decay is given by the form below,
\begin{eqnarray}
\hspace{-1em}\mathcal{P}^{({\rm osc})}_p(t)\sim 
2\mathfrak{P}^{({\rm osc})}_{p,1,2} \left(m_s t\right)^{-2-\alpha_1-\alpha_2}
\cos \left(\frac{\pi}{2}\left(\alpha_2-\alpha_1\right) + \varpi_p t\right), \label{P2MLtosc}
\end{eqnarray}
for $t\gg 1/m_s$. The parameters $\mathfrak{P}^{({\rm osc})}_{p,1,2}$ and $\varpi_p$ are given by the following expressions,
\begin{eqnarray}
&&\hspace{-1em}\mathfrak{P}^{({\rm osc})}_{p,1,2}= \Pi_{j=1}^2 \left|l_j\right|^2 c_{0,j}
\chi_{p,j}^{1+\alpha_j}, \nonumber 
\end{eqnarray}
and
\begin{eqnarray}
&&\hspace{-1em}\varpi_p=\mu_{0,2} \chi_{p,2}-\mu_{0,1} \chi_{p,1}. \label{o21m}
\end{eqnarray}
The frequency of the damped oscillations is the absolute value $\left|\varpi_p\right|$, and coincides with the expression (\ref{o21m}), if $\mu_{0,1}<\mu_{0,2}$, or is the opposite, if $\mu_{0,2}<\mu_{0,1}$. Even if the evolution is not periodic, we naturally define the period $T_p$ of the damped oscillations in term of the frequency of the oscillations as below,
\begin{eqnarray}
T_p=\frac{2 \pi}{\left|\varpi_p\right|}. \label{Tp}
\end{eqnarray}
A similar expression for the frequency of the damped oscillations of the survival probability is obtained in Refs. \cite{HEP_Shir2004,HEP_Shir2006}. In fact, damped oscillations of the survival probability are found in Ref. \cite{HEP_Shir2006} for unstable meson systems in case the two mass states are described by MDDs of different Breit-Wigner forms. The frequency of the oscillations is obtained from the expression (\ref{o21m}) by substituting the lower bounds of the MDDs with the corresponding rest masses. Due to the regular oscillations, the two-mass unstable system is proposed as a quantum clock. See Ref. \cite{HEP_Shir2004} for details.

If the lower bounds of the two mass spectra coincide, $\mu_{0,1}=\mu_{0,2}$, the damped long-time oscillations of the survival probability disappear, since the dominant part of the term $\mathcal{P}^{({\rm osc})}_p(t)$ becomes an inverse power law over long times,
\begin{eqnarray}
\hspace{-1em}\mathcal{P}^{({\rm osc})}_p(t)\sim 
2\cos \left(\frac{\pi}{2}\left(\alpha_2-\alpha_1\right) \right)\mathfrak{P}^{({\rm osc})}_{p,1,2} \left(m_s t\right)^{-2-\alpha_1-\alpha_2}
, \label{P2MLtoscNnpl}
\end{eqnarray}
for $t \gg 1/m_s$. The above form is obtained from Eq. (\ref{P2MLtosc}) in case $\mu_{0,1}=\mu_{0,2}$. 

For the sake of completeness, we consider also the case where the MDDs exhibit the same low-mass power-law profile, $\alpha_1=\alpha_2$, and the same lower bound of the mass spectrum, $\mu_{0,1}=\mu_{0,2}$. In this condition, the long-time survival probability exhibits no oscillations and results in the following dominant inverse-power-law decay,
\begin{eqnarray}
\hspace{0em}\mathcal{P}_p(t)\sim \mathfrak{P}_{p,0}
\left(m_s t\right)^{-2 \left(1+\alpha_1\right)},
\label{Pp0}
\end{eqnarray}
for $t \gg 1/m_s$. The parameter $\mathfrak{P}_{p,0}$ reads
\begin{eqnarray}
\hspace{-2em}\mathfrak{P}_{p,0}=\left(\sum_{j=1,2}
\left|l_j\right|^2 c_{0,j}\right)^2
\chi_{p,1}^{2 \left(1+\alpha_1\right)}. \nonumber
\end{eqnarray}

In summary, if the low-mass power-law profiles of the MDDs differ, $\alpha_1\neq\alpha_2$, the survival probability $\mathcal{P}_p(t)$ exhibits the long-time inverse-power-law decay $\tau^{-2\left(1+\alpha_k\right)}$, where $\alpha_k=\min\left\{\alpha_1,\alpha_2\right\}$. The decay is described by Eq. (\ref{P2MLt1}), if $\alpha_1<\alpha_2$, while, if $\alpha_2<\alpha_1$, the decay is obtained from Eq. (\ref{P2MLt1}) by exchanging the indexes $1$ and $2$. If the low-mass power-law profiles of the two MDDs coincide, $\alpha_1=\alpha_2=\alpha$, and the two mass spectra have the same lower bounds, $\mu_{0,1}=\mu_{0,2}$, the survival probability $\mathcal{P}_p(t)$ exhibits the long-time inverse power-law decay $\tau^{-2\left(1+\alpha\right)}$ which is described by Eq. (\ref{Pp0}). If the low-mass power-law profiles of the two MDDs coincide, $\alpha_1=\alpha_2=\alpha$, and the lower bounds of the mass spectra differ, $\mu_{0,1}\neq\mu_{0,2}$, the long-time survival probability $\mathcal{P}_p(t)$ is described by Eqs. (\ref{P2MLt}), (\ref{P2MLt12}) and (\ref{P2MLtosc}), and exhibits damped oscillations which are enveloped in an inverse-power-law profile, $\tau^{-2\left(1+\alpha\right)}\left(\mathfrak{P}^{({\rm p.l.})}_{p,1,2}+2\mathfrak{P}^{({\rm osc})}_{p,1,2}\cos \left( \varpi_p t\right)\right)$.

\begin{figure*}
\includegraphics[width=0.6\textwidth]{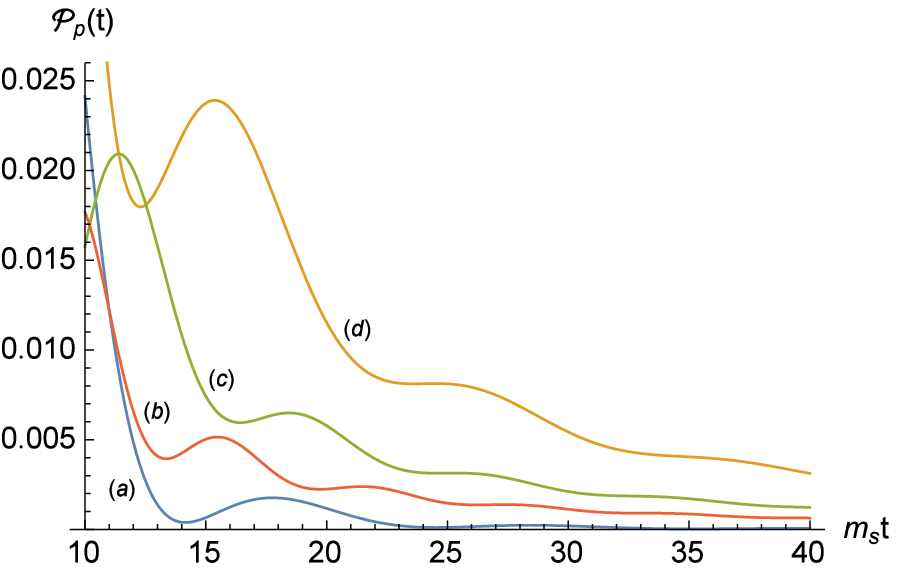}
\caption{(Color online) Survival probability $\mathcal{P}_p(t)$ versus $m_s t$, for 
$ 10 \leq m_s t \leq 40$, and different values of the linear momentum $p$. The computed initial two-mass states are described by Eq. (\ref{2Minitial}) with $l_1=l_2=1/\sqrt{2}$. The corresponding auxiliary functions are given by Eq. 
 (\ref{Ofigs}) with $\xi_{0,1}=1$ and $\xi_{0,2}=2$, for different values of the powers $\alpha_1$ and $\alpha_2$. Curve $(a)$ corresponds to $\alpha_1=1$, $\alpha_2=2$ and $p= 2 m_s$. Curve $(b)$ corresponds to $\alpha_1=0$, $\alpha_2=2$ and $p=  0 m_s$. Curve $(c)$ corresponds to $\alpha_1=0$, $\alpha_2=2$ and $p=  m_s$. Curve $(d)$ corresponds to $\alpha_1=0$, $\alpha_2=2$ and $p= 2 m_s$. 
}
\label{fig5}
\end{figure*}

\begin{figure*}
\includegraphics[width=0.6\textwidth]{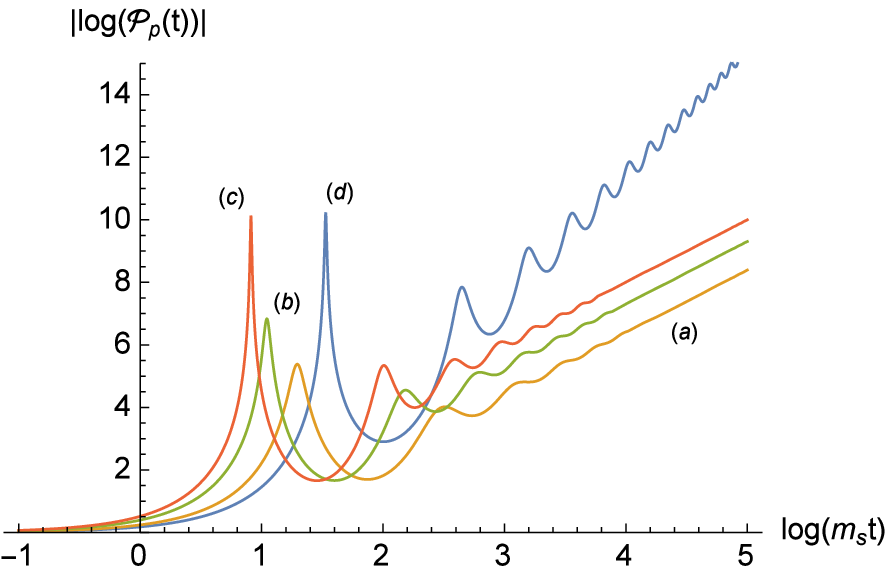}
\caption{(Color online) Quantity $\left|\log\left(\mathcal{P}_p(t)\right)\right|$ versus $\log\left(m_s t\right)$, for 
$ e^{-1} \leq m_s t \leq e^{5}$, and different values of the linear momentum $p$. The computed initial two-mass states are described by Eq. (\ref{2Minitial}) with $l_1=l_2=1/\sqrt{2}$. The corresponding auxiliary functions are given by Eq. (\ref{Ofigs}) with $\xi_{0,1}=1$ and $\xi_{0,2}=2$, for different values of the powers $\alpha_1$ and $\alpha_2$. Curve $(a)$ corresponds to $\alpha_1=0$, $\alpha_2=2$ and $p= 2 m_s$. Curve $(b)$ corresponds to $\alpha_1=0$, $\alpha_2=2$ and $p= m_s$. Curve $(c)$ corresponds to $\alpha_1=0$, $\alpha_2=2$ and $p= 0 m_s$. Curve $(d)$ corresponds to $\alpha_1=1$, $\alpha_2=2$ and $p= 2 m_s$. 
}
\label{fig6}
\end{figure*}

\begin{figure*}
\includegraphics[width=0.6\textwidth]{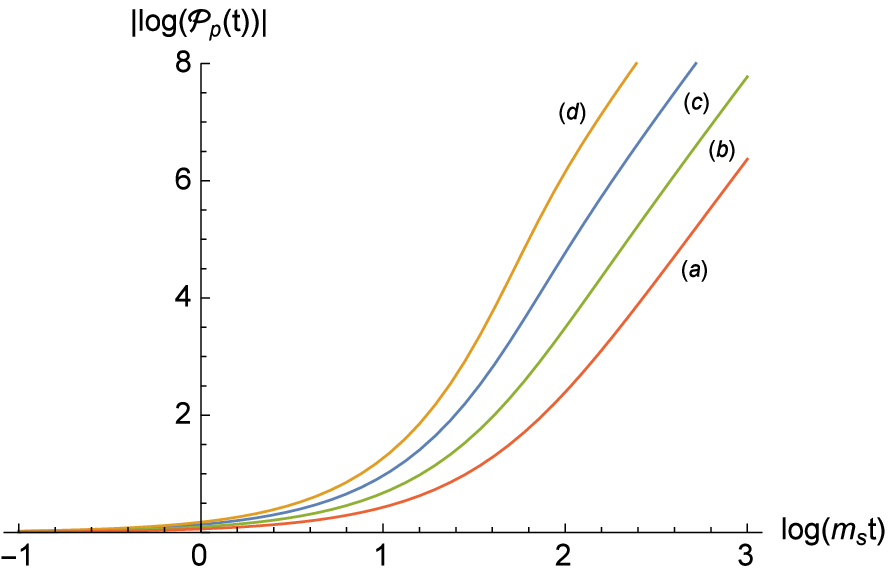}
\caption{(Color online) Quantity $\left|\log\left(\mathcal{P}_p(t)\right)\right|$ versus $\log\left(m_s t\right)$, for 
$ e^{-1} \leq m_s t \leq e^{3}$, and different values of the linear momentum $p$. The computed initial two-mass states are described by Eq. (\ref{2Minitial}) with $l_1=l_2=1/\sqrt{2}$. The corresponding auxiliary functions are given by Eq. (\ref{Ofigs}) for common values of the parameters $\xi_{0,1}$ and $\xi_{0,2}$ and different values of the powers, $\alpha_1=1$ and $\alpha_2=2$. Curve $(a)$ corresponds to $\xi_{0,1}=\xi_{0,2}=2$ and $p= 2 m_s$. Curve $(b)$ corresponds to $\xi_{0,1}=\xi_{0,2}=2$ and $p= 0m_s$. Curve $(c)$ corresponds to $\xi_{0,1}=\xi_{0,2}=1$, 
$\alpha_1=1$, $\alpha_2=2$ and $p=  m_s$. Curve $(d)$ corresponds to $\xi_{0,1}=\xi_{0,2}=1$ and $p= 0 m_s$. 
}
\label{fig7}
\end{figure*}

Numerical analysis of the survival probability $\mathcal{P}_p(t)$ is displayed in Figures \ref{fig1}, \ref{fig2}, \ref{fig3}, \ref{fig4}, \ref{fig5}, \ref{fig6}, \ref{fig7} and \ref{fig8}. The computed initial two-mass states are described by Eq. (\ref{2Minitial}) with $l_1=l_2=2^{-1/2}$, and by toy MDDs which are given by the following auxiliary functions,
\begin{eqnarray}
\Omega_j\left(\xi\right)= w_{j} \xi \left(\xi^2-\xi^2_{0,j}\right)^{\alpha_j} e^{- \xi^2},\label{OmegaFigj}
 \label{Ofigs}
\end{eqnarray}
for every $j=1,2$. 
The parameters $w_{1}$ and $w_2$ are normalization factors. The computation has been performed by considering various values of the parameters $\xi_{0,1}$ and $\xi_{0,2}$, of the non-negative powers $\alpha_1$ and $\alpha_2$, and of the linear momentum $p$. The presence of long-time oscillations in Figures \ref{fig1} and \ref{fig5}, which are computed for $\mu_{0,1}\neq\mu_{0,2}$, and the lack of long-time oscillations in Figure \ref{fig2}, which are computed for $\mu_{0,1}=\mu_{0,2}$, agree with the asymptotic forms of the survival probability which are described by Eqs. (\ref{P2MLt})-(\ref{Tp}). The short-time linear growth of the curves which are displayed in Figures \ref{fig3} and \ref{fig4}, is in accordance with the short-time algebraic decay of the survival probability, which is given by Eq. (\ref{Pptshort2M}). The oscillating behavior and the asymptotic lines appearing in the log-log plot of Figure \ref{fig6}, agree with the oscillatory asymptotic decay of the survival probability which is given by Eqs. (\ref{P2MLt}) and (\ref{P2MLt12})-(\ref{Tp}). The asymptotic lines which appear in the log-log plot of Figure \ref{fig7}, are in accordance with the inverse-power-law decay of the survival probability which is given by Eqs. (\ref{P2MLt}) and (\ref{P2MLt1}). The undamped oscillations which are displayed in Figure \ref{fig8}, agree with the long-time oscillatory decay of the survival probability which is given by Eqs. (\ref{P2MLt}) and (\ref{P2MLt12})-(\ref{Tp}). The oscillations are in accordance with the expression (\ref{Tp}) of the period.

\subsection{Nonrelativistic and ultrarelativistic limits}\label{32}

The long-time behavior of the survival probability
in the nonrelativistic limit or, equivalently, in the rest reference frame of the moving unstable system, is obtained from Eqs. (\ref{P2MLt})-(\ref{Pp0}), for $p=0$.
The survival probability $\mathcal{P}_0(t)$ is approximated over long times, $t \gg 1/m_s$, by the sum of the inverse-power-law term $\mathcal{P}^{({\rm p.l})}_p(t)$ and by the damped oscillating term $\mathcal{P}^{({\rm osc})}_0(t)$,
\begin{eqnarray}
\mathcal{P}_0(t)\sim \mathcal{P}^{({\rm p.l})}_0(t)+
\mathcal{P}^{({\rm osc})}_0(t), \label{P2MLtp0}
\end{eqnarray}
for $t \gg 1/\omega_s$. If the low-mass power-law profiles differ 
and $\alpha_1<\alpha_2$ we find 
\begin{eqnarray}
\mathcal{P}^{({\rm p.l.})}_0(t)\sim 
\mathfrak{P}^{({\rm p.l.})}_{0,1}\left(m_s t\right)^{-2\left(1+\alpha_1\right)},
\label{P2MLt10}
\end{eqnarray}
for $t \gg /m_2$;
while, if $\alpha_1=\alpha_2$, we obtain
\begin{eqnarray}
\mathcal{P}^{({\rm p.l.})}_0(t)\sim \mathfrak{P}^{({\rm p.l.})}_{0,1,2}
\left(m_s t\right)^{-2\left(1+\alpha_1\right)}, 
\label{P2MLt12p0}
\end{eqnarray}
for $t \gg /m_2$. The coefficients $\mathfrak{P}^{({\rm p.l.})}_{0,1}$ and $\mathfrak{P}^{({\rm p.l.})}_{0,1,2}$ are given by the expression below,
\begin{eqnarray}
&&\hspace{-3em}\mathfrak{P}^{({\rm p.l.})}_{0,1}=
\left|l_1 \right|^4 c_{0,1}^2, 
\hspace{1em}\mathfrak{P}^{({\rm p.l.})}_{0,1,2}=\sum_{j=1}^2
\left|l_j \right|^4 c_{0,j}^2 . \nonumber
\end{eqnarray}
The oscillating term $\mathcal{P}^{({\rm osc})}_0(t)$ reads 
\begin{eqnarray}
\hspace{-1em}\mathcal{P}^{({\rm osc})}_0(t)\sim 
2\mathfrak{P}^{({\rm osc})}_{0,1,2} \left(m_s t\right)^{-2-\alpha_1-\alpha_2}
\cos \left(\frac{\pi}{2}\left(\alpha_2-\alpha_1\right) + \varpi_0 t\right), \label{P2MLtoscNnplp0}
\end{eqnarray}
for $t\gg 1/m_s$, if $\alpha_1\leq\alpha_2$. The parameters $\mathfrak{P}^{({\rm osc})}_{0,1,2}$ and $\varpi_0$ are given by the expressions below,
\begin{eqnarray}
&&\hspace{-1em}\mathfrak{P}^{({\rm osc})}_{0,1,2}= \Pi_{j=1}^2 \left|l_j\right|^2 c_{0,j}, \nonumber 
\end{eqnarray}
and
\begin{eqnarray}
&&\hspace{-1em}\varpi_0=\mu_{0,2}-\mu_{0,1} . \label{o21mp0}
\end{eqnarray}
The frequency of the damped oscillations is $\left|\mu_{0,2}-\mu_{0,1}\right|$, and the period $T_0$ of the damped oscillations is
\begin{eqnarray}
T_0=\frac{2 \pi}{\left|\mu_{0,2}-\mu_{0,1}\right|}. \label{T0}
\end{eqnarray}
If the lower bounds of the two mass spectra coincide, $\mu_{0,1}=\mu_{0,2}$, the term $\mathcal{P}^{({\rm osc})}_0(t)$ results in a dominant inverse power law over long times,
\begin{eqnarray}
\hspace{-1em}\mathcal{P}^{({\rm osc})}_0(t)\sim 
2\cos \left(\frac{\pi}{2}\left(\alpha_2-\alpha_1\right) \right)\mathfrak{P}^{({\rm osc})}_{0,1,2} \left(m_s t\right)^{-2-\alpha_1-\alpha_2}
, \label{P2MLtoscNn0}
\end{eqnarray}
for $t \gg 1/m_s$.
If the MDDs exhibit the same low-mass power-law profile, $\alpha_1=\alpha_2$, and 
the same lower bound of the mass spectrum, $\mu_{0,1}=\mu_{0,2}$, the long-time survival probability reads 
\begin{eqnarray}
\hspace{-1em}\mathcal{P}_0(t)\sim \mathfrak{P}_{0,0}
\left(m_s t\right)^{-2 \left(1+\alpha_1\right)}
\label{P00}
\end{eqnarray}
for $t \gg 1/m_s$, where 
$$\mathfrak{P}_{0,0}=\left(\sum_{j=1,2}
\left|l_j\right|^2 c_{0,j}\right)^2
.$$

In the ultrarelativistic limit, 
$p \gg \max\left\{\mu_{0,1},\mu_{0,2}\right\}$, the long-time survival probability is still given by Eqs. (\ref{P2MLt})-(\ref{Pp0}), but the involved factors are approximated by the following forms,
\begin{eqnarray}
&&\hspace{-3em}\mathfrak{P}^{({\rm p.l.})}_{p,1}\simeq
\left|l_1 \right|^4 c_{0,1}^2 \left(\frac{p}{\mu_{0,1}}\right)^{2\left(1+\alpha_1\right)}, %\nonumber\\&&\hspace{-1em}
\hspace{1em}\mathfrak{P}^{({\rm p.l.})}_{p,1,2}\simeq
\sum_{j=1}^2
\left|l_j \right|^4 c_{0,j}^2 \left(\frac{p}{\mu_{0,j}}\right)^{2 \left(1+\alpha_j\right)}, \nonumber \\
&&\hspace{-3em}\mathfrak{P}^{({\rm osc})}_{p,1,2}\simeq p^{2+\alpha_1+\alpha_2}\Pi_{j=1}^2 \left|l_j\right|^2 c_{0,j}
\mu_{0,j}^{-\left(1+\alpha_j\right)}, \hspace{1em}
\mathfrak{P}_{p,0}\simeq\left(\sum_{j=1,2}
\left|l_j\right|^2 c_{0,j}\right)^2
\left(\frac{p}{\mu_{0,1}}\right)^{2 \left(1+\alpha_1\right)}. \nonumber 
\end{eqnarray}
The frequency of the oscillations $\left|\varpi_p\right|$ vanishes and, consequently, the period $T_p$ diverges in the ultrarelativistic limit.

\section{Time dilation in the decay laws}\label{4}

For a moving single-mass unstable quantum state the survival probability at rest $\mathcal{P}_0(t)$ transforms in the survival probability $\mathcal{P}_p(t)$, approximately, according to a scaling law, over long times \cite{Gxiv2018}. Similar scaling relation is found, under certain conditions, in the present transformation of the decay laws of moving two-mass unstable quantum states. In fact, if the low-mass power-law profiles of the MDDs $\omega_1\left(m\right)$ and $\omega_2\left(m\right)$ differ, $\alpha_1\neq\alpha_2$, the survival probability $\mathcal{P}_p(t)$, which is given by Eq. (\ref{P2MLt}), is described by the term $\mathcal{P}^{({\rm p.l.})}_p(t)$, over long times. %This term is given by Eq. (\ref{P2MLt1}) and results in a dominant long-time inverse-power-law decay. 
In this case, the survival probability at rest $\mathcal{P}_0(t)$ transforms in the survival probability $\mathcal{P}_p(t)$ according to the following scaling law over long times,
\begin{eqnarray}
\mathcal{P}_p(t)\sim \mathcal{P}_0\left(\frac{t}{\chi_{p,k}}\right),
\label{Pdilation}
\end{eqnarray}
for $t \gg 1/m_s$. The scaling factor $\chi_{p,k}$ is given by the expression below,
\begin{eqnarray}
\chi_{p,k}=\sqrt{1+\frac{p^2}{\mu_{0,k}^2}}. \label{chipk}
\end{eqnarray}
The index $k$ takes the value $1$ or $2$ and is selected by the lower of the powers $\alpha_1$ and $\alpha_2$. 
% as follows, $\alpha_k = \min \left\{\alpha_1,\alpha_2\right\}$. 
The long-time scaling transformation (\ref{Pdilation}), which is induced by the change of reference frame, represents a dilation of times with scaling factor $\chi_{p,k}$. This scaling factor is determined by the linear momentum $p$ of the moving unstable system and by the lower bound $\mu_{0,k}$ of the MDD.

 Similarly to the case of a moving single-mass unstable quantum system \cite{Gxiv2018}, the scaling factor $\chi_{p,k}$ coincides with the relativistic Lorentz factor of a mass at rest $\mu_{0,k}$ which moves with linear momentum $p$. Consequently, also for the moving two-mass unstable quantum states under study, the scaling transformation of the survival probability can be interpreted as the effect of the relativistic time dilation. This interpretation holds if the lower bound $\mu_{0,k}$ is considered to be the effective mass at rest of the unstable system over long times, since the unstable system moves with linear momentum $p$. This interpretation suggests the value $1/\sqrt{1+\mu_{0,k}^2/p^2}$ as the constant asymptotic velocity of the moving two-mass system.

For the sake of completeness, we consider also a further situation. Let $\mathcal{P}_{p^{\prime}}(t)$ be the survival probability which is detected in the reference frame where the two-mass unstable system moves with constant linear momentum $p^{\prime}$. Relation (\ref{Pdilation}) suggests that the survival probability $\mathcal{P}_{p}(t)$ is linked to the survival probability $\mathcal{P}_{p^{\prime}}(t)$, over long times, by the following scaling transformation,
\begin{eqnarray}
\mathcal{P}_p(t)\sim \mathcal{P}_{p^{\prime}}\left(\frac{\chi_{p^{\prime},k}}{\chi_{p,k}}t\right),
\label{PpPpprime}
\end{eqnarray}
for $t \gg 1/m_s$. The corresponding scaling factor is the ratio $\chi_{p,k}/\chi_{p^{\prime},k}$. If the linear momentum $p$ is greater than $p^{\prime}$, i.e., $0\leq p^{\prime}<p$, the survival probability $\mathcal{P}_{p^{\prime}}\left(t\right)$ transforms in the survival probability $\mathcal{P}_p(t)$ according to a time dilation, since the scaling factor $\chi_{p,k}/\chi_{p^{\prime},k}$ is greater than unity.

\begin{figure*}
\includegraphics[width=0.6\textwidth]{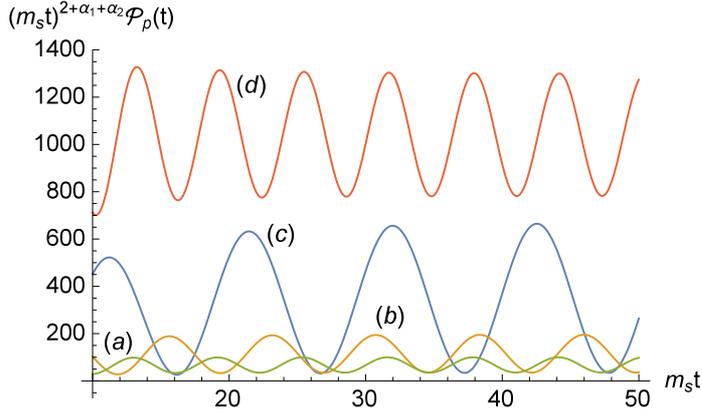}
\caption{(Color online) Quantity $\left(m_s t\right)^{2+\alpha_1+\alpha_2}\mathcal{P}_p(t)$ versus $m_s t$, for 
$ 10 \leq m_s t \leq 50$, and different values of the linear momentum $p$. The computed initial two-mass states are described by Eq. (\ref{2Minitial}) with $l_1=l_2=1/\sqrt{2}$. The corresponding auxiliary functions are given by Eq. (\ref{Ofigs}) with $\xi_{0,1}=1$ and $\xi_{0,2}=2$, for common values of the powers $\alpha_1$ and $\alpha_2$. Curve $(a)$ corresponds to  $\alpha_1=\alpha_2=1$ and $p= 0 m_s$. Curve $(b)$ corresponds to $\alpha_1=\alpha_2=1$ and $p=  m_s$. Curve $(c)$ corresponds to $\alpha_1=\alpha_2=1$ and $p= 2 m_s$. Curve $(d)$ corresponds to $\alpha_1=\alpha_2=2$ and $p= 0 m_s$. 
}
\label{fig8}
\end{figure*}

\begin{figure*}
\includegraphics[width=0.6\textwidth]{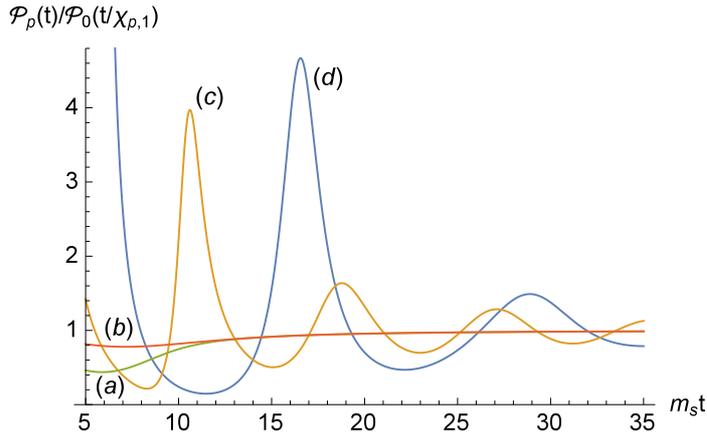}
    \caption{ \small (Color online) Ratio $\mathcal{P}_p\left(t\right)/\mathcal{P}_0\left(t/\chi_{p,1}\right)$ for $5\leq m_s t \leq 35$, and different values of the linear momentum $p$. The computed initial two-mass states are described by Eq. (\ref{2Minitial}) with $l_1=l_2=1/\sqrt{2}$. The corresponding auxiliary functions are given by Eq. (\ref{Ofigs}) with $\alpha_1=1$ and $\alpha_2=2$, for various values of the parameters $\xi_{0,1}$ and $\xi_{0,2}$. Curve $(a)$ corresponds to $\xi_{0,1}=\xi_{0,2}=1$, and $p= m_s$. Curve $(b)$ corresponds to $\xi_{0,1}=\xi_{0,2}=2$, and $p= 2 m_s$. Curve $(c)$ corresponds to $\xi_{0,1}=1$, $\xi_{0,2}=2$, and $p=m_s$. Curve $(d)$ corresponds to $\xi_{0,1}=1$, $\xi_{0,2}=2$, and $p=m_s$. 
		}
		\label{fig9}
\end{figure*}

In Figure \ref{fig9}, each curve tends to the same asymptotic horizontal line, at ordinate $1$, with either oscillatory or monotone behavior. This asymptotic condition confirms the scaling transformation (\ref{Pdilation}) of the survival probability and, consequently, the dilation of times which occurs by changing reference frame. Notice that the scaling law holds if the low-mass power-law profiles of the MDDs differ, $\alpha_1 \neq \alpha_2$, both in presence, $\mu_{0,1}\neq\mu_{0,2}$, and in absence, $\mu_{0,1}=\mu_{0,2}$, of damped oscillations of the survival probability.

The long-time scaling transformation (\ref{Pdilation}) of the survival probability is lost if the MDDs exhibit the same power-law profiles, $\alpha_1=\alpha_2$, and the lower bounds of the mass spectra differ, $\mu_{0,1}\neq\mu_{0,2}$. Under this condition the long-time survival probability $\mathcal{P}_p(t)$ is described by Eqs. (\ref{P2MLt}), (\ref{P2MLt12}) and (\ref{P2MLtosc}), and exhibits the damped oscillations which are given by Eq. (\ref{P2MLtosc}). The period $T_p$ of the long-time damped oscillations is given by Eq. $(\ref{Tp})$ and holds for every nonnegative value $p$ of the linear momentum. Consequently, the ratio between the period $T_p$ and the period at rest $T_0$ shows how the oscillations transform,
\begin{eqnarray}
\frac{T_p}{T_0}=\frac{\varpi_0}{\varpi_p}=\sum_{j=0}^1\frac{\mu_{0,j}}{\mu_{0,1}+\mu_{0,2}} \chi_{p,j}. \label{TpT0}
\end{eqnarray}
The above relation suggests that the long-time oscillations of the survival probability are dilated by the change of the reference frame. %a dilatation appears in the period of  . 
In fact, the period $T_0$ of the oscillations at rest transforms in the reference frame where the unstable system moves with linear momentum $p$, according to a factor which is larger than unity. This factor is the weighted mean of the scaling factors $\chi_{p,1}$ and $\chi_{p,2}$. The non-normalized weights are the lower bounds of the mass spectra. Similar transformation of the frequency of the damped oscillations is obtained from the detailed analysis which is performed in Ref. \cite{HEP_Shir2004}. The period of those oscillations transforms according to the relativistic dilation of times if the rest masses, appearing in the Breit-Wigner forms of the MDDs, are approximately equal. See Ref. \cite{HEP_Shir2004} for details. %Notice that the frequency of the present oscillations are recovered from those which are described in Ref. \cite{HEP_Shir2004} by substituting the rest mass of each state with the corresponding lower bound of the mass spectrum.

For the sake of completeness, consider the decay which is observed in the reference frame where the unstable system moves with constant linear momentum $p^{\prime}$. If $p> p^{\prime}\geq 0$, relation (\ref{TpT0}) suggests that the period $T_p$ of the oscillations is enlarged with respect to the period $T_{p^{\prime}}$, according to the following factor,
\begin{eqnarray}
\frac{T_p}{T_{p^{\prime}}}=\frac{\varpi_{p^{\prime}}}{\varpi_p}=\frac{\sum_{j=0}^1\mu_{0,j} \chi_{p,j}}{\sum_{j=0}^1 \mu_{0,j} \chi_{p^{\prime},j}}. \label{TpTpprime}
\end{eqnarray}

The periodic oscillations appearing in Figure \ref{fig8}, show a dilation in the period of the damped oscillations of the survival probability which occurs by changing the reference frame. The magnitudes of the dilated periods are in accordance with the factor which is given by Eq. (\ref{TpT0}).

\section{Summary and conclusions}\label{5}

We have considered an unstable quantum state which is initially prepared in a superposition of two mass states. The decay is studied in a laboratory reference frame where the unstable system is moving with constant linear momentum $p$. Each mass state is an eigenstate of the linear momentum to the common eigenvalue $p$. The evolution of each mass state is assumed to be approximately orthogonal to the other mass state. This assumption is based upon the almost exact CP-conservation in the decay of unstable meson systems \cite{HEP_Shir2004}. The two mass states are described by MDDs which are different from each other and are tailored as power laws, with powers $\alpha_1$ and $\alpha_2$, near the non-vanishing lower bounds, $\mu_{0,1}$ and $\mu_{0,2}$, respectively, of the mass spectra. The MDDs are arbitrarily tailored over higher values of the mass variable, except for some conditions which involve the orthogonality of the two mass states, and the differentiability and integrability of the MDDs. 

The survival probability $\mathcal{P}_p(t)$ has been analyzed over short and long times in the reference frame where the unstable system moves with constant linear momentum $p$. Due to the generality of the linear momentum, the ultrarelativistic and non-relativistic limits have been evaluated as particular cases. 
Over short times, the survival probability $\mathcal{P}_p(t)$ decays algebraically. Over long times, the survival probability exhibits a dominant inverse-power-law decay or damped oscillations which are enveloped in an inverse-power-law profile. The appearance of each regime is determined by the low-mass properties of the two MDDs.

 If the powers $\alpha_1$ and $\alpha_2$ differ, the dominant inverse-power-law decay $\tau^{-2(1+\alpha_k)}$ manifests over long times. The power $\alpha_k$ is the minimum between the powers $\alpha_1$ and $\alpha_2$. In this case the long-time survival probability $\mathcal{P}_p(t)$ is approximately related to the survival probability at rest $\mathcal{P}_0(t)$ by the following scaling law, $\mathcal{P}_p(t)\sim \mathcal{P}_0 \left(t/\chi_{p,k}\right)$. This scaling property suggests that the long-time survival probability at rest transforms in the laboratory frame where the unstable system moves with linear momentum $p$, according to a dilation of times. The corresponding scaling factor $\chi_{p,k}$ is determined by the MDD with the lower power-law profile and reads $\sqrt{1+p^2/\mu_{0,k}^2}$. Consequently, the transformation of the long-time survival probability consists in a time dilation if the two MDDs are tailored with different power laws near the lower bound of the corresponding mass spectrum. The time dilation appears independently of the (non-vanishing) values of the lower bounds. %The time dilation of the two-mass decay coincides with the time dilation which appears in the single-mass decay of the unstable state \cite{Gxiv2018}, which exhibits the lower power-law profile.
% The long-time survival probability transforms according to . The corresponding single-mass unstable state is described by the MDD with the lower power-law profile. 
The scaling factor $\chi_{p,k}$ of the time dilation coincides with the relativistic Lorentz factor of a mass at rest $\mu_{0,k}$ which moves with linear momentum $p$. This observation suggests the following interpretation of the transformation of the decay laws in terms of the theory of special relativity. The present time dilation reproduces the relativistic dilation of times if the lower bound $\mu_{0,k}$ of the mass spectrum is accounted as the mass at rest of the unstable system, which moves with constant linear momentum $p$. 

If the powers $\alpha_1$ and $\alpha_2$ coincide and the lower bounds $\mu_{0,1}$ and $\mu_{0,2}$ of the mass spectra differ, the survival probability $\mathcal{P}_p(t)$ decays over long times according to damped oscillations. The oscillations are enveloped in the inverse-power-law profile $\tau^{-2(1+\alpha)}$, where $\alpha$ is the common value of the two powers. The frequency of the oscillations is determined by the lower bounds $\mu_{0,1}$ and $\mu_{0,2}$ and by the linear momentum $p$, and reads $\left|\mu_{0,2} \chi_{p,2}-\mu_{0,1} \chi_{p,1}\right|$. If the powers $\alpha_1$ and $\alpha_2$ coincide, the long-time scaling relation which links the survival probability $\mathcal{P}_p$ and the survival probability at rest $\mathcal{P}_0(t)$, is lost. The period $T_p$ of the oscillations dilates with respect to the period $T_0$ of the oscillations at rest. Equivalently, the frequency $\left|\varpi_p\right|$ diminishes with respect to the frequency at rest $\left|\varpi_0\right|$. The factor $T_p/T_0$, or, equivalently, $\left|\varpi_0/\varpi_p\right|$, consists in the weighted average of the scaling factors $\chi_{p,1}$ and $\chi_{p,2}$. The non-normalized weights are the lower bounds $\mu_{0,1}$ and $\mu_{0,2}$ of the mass spectra.

In conclusion, we have found decay laws of a moving two-mass unstable quantum state which consist in dominant inverse power laws or in damped oscillations, over long times. The appearance of each regime is determined by the low-mass differences between the MDDs. In the long-time inverse-power-law regime the transformation of the decay laws, which is induced by the change of the reference frame, consists in a time dilation. In the long-time damped oscillatory regime the time dilation is lost. Still, the period of the oscillations transforms regularly. In both these regimes the properties of the transformed long-time decays are determined by the (constant) linear momentum of the moving unstable system and by the (non-vanishing) lower bounds of the mass spectra.

\appendix\label{A}
\section{Details}

The short- and long-time behavior of the survival probability of a moving two-mass unstable quantum system are evaluated from the asymptotic forms of the survival amplitude of a single-mass unstable quantum system which moves with constant linear momentum $p$ in the laboratory frame of an observer. These asymptotic forms are reported below, for the sake of clarity, by following Ref. \cite{Gxiv2018}. 

If the auxiliary function of the MDD of a single-mass unstable quantum system vanishes as $\Omega\left(\xi \right)= \mathcal{O}\left(\xi^{-1-l_0}\right)$ for $\xi\to+\infty$, with $l_0>5$, the survival amplitude decays algebraically, approximately, over short times, 
\begin{eqnarray}
A_p(t)\sim 1-\imath a_0 t-a_1 t^2+ \imath a_2 t^3, \label{Aptshort}
\end{eqnarray}
for $t \ll 1/ m_s$. The constants $a_0$, $a_1$ and $a_2$ read
\begin{eqnarray}
 &&\hspace{-3em}a_0=\int_{\mu_0}^{\infty}\omega\left(m\right)\sqrt{p^2+m^2}  dm,
\hspace{1em}a_1=\frac{1}{2}\int_{\mu_0}^{\infty}\omega\left(m\right) \left(p^2+m^2\right)  dm, \nonumber
 \\&&\hspace{-3em}a_2=\frac{1}{6}\int_{\mu_0}^{\infty}\omega\left(m\right)\left(p^2+m^2\right)^{3/2}  dm.
\nonumber 
\end{eqnarray}
The above asymptotic form of the survival amplitude provides the short-time behavior of the survival probability, which is given by Eq. (\ref{Pptshort}), with $\pi_0=2 a_1-a_0^2$.

 Over long times the survival amplitude of the moving single-mass unstable quantum system results in the following expression \cite{Gxiv2018},
\begin{eqnarray}
A_p(t)\sim c_0  e^{-\imath \left(\left(\pi/2\right)
\left(1+\alpha\right)+\sqrt{\mu_0^2+p^2} t\right)}
\left(\frac{\chi_p}{m_s t}\right)^{1+\alpha},
\label{Aplongt}
\end{eqnarray}
for $t \gg 1/ m_s$, where $c_0=\Gamma\left(1+\alpha\right)
\Omega_0\left(\xi_0\right)$. The factor $\chi_p$ is given by Eq. (\ref{Chip}). The asymptotic form (\ref{Aplongt}) provides the long-time expressions of the survival probability $\mathcal{P}_p(t)$ and of the survival probability at rest $\mathcal{P}_0(t)$, which are given by the asymptotic forms (\ref{Pplongt}) and (\ref{P0longt}), respectively. The comparisons of these expressions provides the long-time scaling property (\ref{PpP0L}).

In case the system is initially prepared in the two-mass unstable quantum state (\ref{2Minitial}), the survival amplitude is given by Eq. (\ref{Ap12}), due to the approximate vanishing of the cross terms 
$\langle\phi_1\left|e^{-\imath H t}\right|\phi_2 \rangle$ and $\langle\phi_2\left|e^{-\imath H t}\right|\phi_1 \rangle$. If the MDDs which describe the mass states $|\phi_1\rangle$ and $|\phi_2\rangle$, belong to the class which is introduced in Section \ref{21}, the short-time behavior of the terms $A_{p,1}(t)$ and $A_{p,2}(t)$ is evaluated via Eq. (\ref{Aptshort}). The survival probability of the moving two-mass unstable state is obtained as the square modulus of the right hand side of Eq. (\ref{Ap12}),
\begin{eqnarray}
\mathcal{P}_p(t)=\sum_{j=1}^{2} \left|l_j\right|^4 \left|A_{p,j}(t)\right|^2+2 
\left|l_1\right|^2\left|l_2\right|^2 \mathrm{ Re}\left\{A_{p,1}(t)A^{\ast}_{p,2}(t)\right\}.
\label{PpAp}
\end{eqnarray}
The short-time expression of the survival probability is found from Eqs. (\ref{Aptshort}) and (\ref{PpAp}), and results in the asymptotic form (\ref{Pptshort2M}).

Over long times, the survival probability is 
obtained from Eq. (\ref{PpAp}) and from the long-time forms of the terms $A_{p,1}(t)$ and $A_{p,2}(t)$, which are evaluated via the asymptotic expression (\ref{Aplongt}). In this way, we find that the long-time behavior of the survival probability is given by 
the term $\left|A_{p,1}(t)\right|^2$, if $\alpha_1<\alpha_2$, and by the term $\left|A_{p,2}(t)\right|^2$, if $\alpha_2<\alpha_1$, and Eqs. (\ref{P2MLt1}) and (\ref{Chipj}) are obtained. If $\alpha_1=\alpha_2$, every term of right hand side of Eq. (\ref{PpAp}) contributes to the dominant part of the asymptotic expansion of the survival probability, over long times. The damped oscillations are generated by the last term which appears in the right hand side of Eq. (\ref{PpAp}), if $\mu_{0,1}\neq\mu_{0,2}$. This term and the asymptotic form (\ref{Aplongt}) provide Eqs. (\ref{P2MLtosc})-(\ref{Tp}).

The long-time behavior of the survival probability in the non-relativistic regime, or, equivalently, at rest, is given by Eqs. (\ref{P2MLtp0})-(\ref{P00}), and is obtained from Eqs. (\ref{P2MLt})-(\ref{Pp0}), via the relation $\chi_{0,j}=1$, which holds for every $j=1,2$. The relation is obtained from Eq. (\ref{Chipj}) in case $p=0$.
In the ultrarelativistic limit, $p\gg \max \left\{\mu_{0,1}, \mu_{0,2}\right\}$, the survival probability is described over long times by Eqs. (\ref{P2MLt})-(\ref{Pp0}). The 
involved factors, $\mathfrak{P}^{({\rm p.l.})}_{p,1}$, $\mathfrak{P}^{({\rm p.l.})}_{p,1,2}$, $\mathfrak{P}^{({\rm osc})}_{p,1,2}$, and $\mathfrak{P}_{p,0}$, are reported in Section \ref{32}. These factors are obtained from the exact forms, which are given in Section \ref{31}, via the following approximations, $\chi_{p,1}\simeq p/\mu_{0,1}$, and $\chi_{p,2}\simeq p/\mu_{0,2}$. These approximations hold for $p\gg \max \left\{\mu_{0,1}, \mu_{0,2}\right\}$.

The long-time scaling relation (\ref{Pdilation}) is found by comparing Eq. (\ref{P2MLt1}) with the form which is obtained from Eq. (\ref{P2MLt1}) for $p=0$. The long-time relation (\ref{PpPpprime}) is obtained from the scaling property (\ref{Pdilation}) by considering the values $p$ and $p^{\prime}$ of the linear momentum. If $\alpha_1=\alpha_2$, the scaling relation (\ref{Pdilation}) fails due to the presence of the oscillatory term and of the coefficients $\mathfrak{P}^{({\rm p.l.})}_{p,1,2}$ and $\mathfrak{P}^{({\rm osc})}_{p,1,2}$. The ratio (\ref{TpT0}) is obtained from Eq. (\ref{Tp}). This concludes the demonstration of the present results.

\end{document}